\documentclass{nature}
\usepackage{graphicx}
\usepackage{dcolumn}
\usepackage{bm}
\usepackage{pstricks}
\usepackage{epsfig}

\title{Longitudinal wave function control in single quantum dots with an applied magnetic field }

\author{Shuo Cao$^1$, Jing Tang$^{1,2}$, Yunan Gao$^1$, Yue Sun$^1$, Kangsheng Qiu$^1$, Yanhui Zhao$^1$, Min He$^1$, Jin-An Shi$^1$, Lin Gu$^1$, David A. Williams$^3$, Weidong Sheng$^4$, Kuijuan Jin$^{1,5,\ast}$ \& Xiulai Xu$^{1,\ast}$}

\begin{document}

\maketitle

\begin{affiliations}
 \item Beijing National Laboratory for Condensed Matter Physics, Institute of Physics, Chinese Academy of Sciences, Beijing, 100190, China
 \item Institute of Photo-electronic Thin Film Devices and Technology, Nankai University, Tianjin, 300071, China
 \item Hitachi Cambridge Laboratory, Cavendish Laboratory, Cambridge CB3 0HE, United Kingdom
 \item Department of Physics, Fudan University, Shanghai, 200433, China
 \item Collaborative Innovation Center of Quantum Matter, Beijing, China

  $^\ast$  Correspondence and requests for materials
should be addressed to K.J.(kjjin@iphy.ac.cn) or X.X.(xlxu@iphy.ac.cn)
\end{affiliations}

 \begin{abstract}
Controlling single-particle wave functions in single semiconductor quantum dots is in demand to implement solid-state quantum information processing and spintronics. Normally, particle wave functions can be tuned transversely by an perpendicular magnetic field. We report a longitudinal wave function control in single quantum dots with a magnetic field. For a pure InAs quantum dot with a shape of pyramid or truncated pyramid, the hole wave function always occupies the base because of the less confinement at base, which induces a permanent dipole oriented from base to apex. With applying magnetic field along the base-apex direction, the hole wave function shrinks in the base plane. Because of the linear changing of the confinement for hole wave function from base to apex, the center of effective mass moves up during shrinking process. Due to the uniform confine potential for electrons, the center of effective mass of electrons does not move much, which results in a permanent dipole moment change and an inverted electron-hole alignment along the magnetic field direction. Manipulating the wave function longitudinally not only provides an alternative way to control the charge distribution with magnetic field but also a new method to tune electron-hole interaction in single quantum dots.
\end{abstract}

\newpage
\maketitle

Semiconductor quantum dots have been investigated intensively for potential applications in solid-state quantum information processing and spintronics, such as single-photon sources~\cite{Kim1999,
Michler2000,Xu2004,Bennett2005,Xu2007,Aoki2008,Akopian2011,Makhonin2011}, excitonic qubits~\cite{Zrenner2002}, spin
qubits~\cite{Kolodka2007,Mikkelsen2007,Greilich2009,Kim2011,Weiss2013,Warburton2013}, and spin-photon entanglement
interfaces~\cite{Greve2012,Schaibley2013,Webster2014}. Controlling single-particle wave functions is a key step to understand quantum physics and to achieve the applications with single semiconductor quantum dots. To manipulate the wave functions in a single quantum dot, an external magnetic field or an electric field is normally applied. For applying a magnetic field, the particle wave functions can be tuned transversely in the presence of cyclotron energy\cite{Kittle1976}. With a transverse magnetic field, the carrier wave functions of different orbitals in single quantum dots have been mapped using tunneling spectroscopy technique~\cite{Vdovin2000,Bester2007,Millo2001,Patane2010,Rontani2011,Lei2010}. Here, we report a longitudinal wave function control with a magnetic field in single quantum dots for the first time. With applying magnetic field along the base-apex direction of quantum dots, the hole wave function shrinks in the base plane due to cyclotron motion. Because of linear changing of the confinement for hole wave function from the base to the apex in quantum dots, the center of effective mass of hole wave functions moves up along that direction during shrinking process. The center of effective mass of electron wave functions does not shift longitudinally much because of the uniform confining potential for electrons. Manipulating the wave function longitudinally not only provides an alternative way to control the charge distribution with magnetic field but also a new method to tune the electron-hole interaction in single quantum dots.

Due to chemical composition variation and pyramidal shape of self-assemble quantum dots, electron-hole pairs generated optically or electrically have a permanent dipole moment along the growth direction because the wave function distribution of electrons is different from that of holes~\cite{Sheng2001}. For pure InAs quantum dots with a pyramidal or truncated pyramidal shape, the permanent dipole moment is in the direction from base to apex due to linear increase of the confining potential of the valence band for heavy holes ~\cite{Cusack1996}, which provides a good candidate to control the charge wave function longitudinally. The permanent dipole can be examined by measuring the excitonic transition energy change as a function of external electric field across the quantum dots, i.e., the quantum-confined Stark effect~\cite{Barker2000,Jin2004,Mar2011,Mar2013,Xu20081,Sheng2002}. An electric field ($F$) is applied, the ground state transition energy varies quadratically as $E(F)=E_{0}+pF+\beta F^2$, where $E_{0}$ is the transition energy with zero-field, $p$ is the permanent dipole moment and $\beta$ is the polarizability of electron-hole wave functions. The asymmetric Stark shift gives the built-in dipole moment, which reveals the spatial distribution of electron-hole wave functions in a single quantum dot. Instead of pure InAs quantum dots, an inverted electron-hole alignment has been observed experimentally in a few special cases with Ga diffusing~\cite{Fry2000}. However, no permanent dipole reverting has been reported in same single quantum dots. Here, the hole wave function can be tuned from bottom to the top of that of electrons with increasing magnetic field because of the different confining potentials, resulting in an inverted electron-hole alignment longitudinally.

\textbf{Results}

\textbf{Charging state control.}
Figure 1(a) shows PL spectra of a single quantum dot with different bias voltages. With a bias voltage at -0.5 V (as shown in the top panel), four peaks from different charging states can be observed. They are singly positively charged exciton (X$^{+}$), neutral exciton(X$^{0}$), singly negatively charged exciton (X$^{-}$) and doubly negatively charged exciton (X$^{2-}$) states respectively, as labeled in the panel. The band diagram of the device with a bias at -0.5 V is sketched in Figure 1(b). With non-resonant optical pumping, the generated electrons tunnel out more easily with a negative bias voltage. It can be seen that electron tunneling results in the quantum dot to be more positively charged. In contrast, the negatively charged exciton emission dominates the spectrum with bias voltages at 0 V and 0.5 V. Due to the Coulomb interaction, the charging energies for X$^{-}$ and X$^{+}$ are 7 meV and 4 meV respectively. However, the energy separation between X$^{-}$ and X$^{2-}$ is only about 300 $\mu$eV, which is due to the fact that third electron in X$^{2-}$ occupies p orbital and has a weaker Coulomb interaction. The details of the assignment for charging states are discussed before \cite{Tang2014}.

To investigate the permanent dipole of single quantum dots in an $n$-$i$-Schottky diode, the direction of built-in electric field is denoted as positive in this work. A negative bias applied to the Schottky contact increases total electric field across the quantum dots, while positive bias voltage decreases total electric field as sketched in Figure 1(c). The bias voltage required to achieve zero total electric field across the quantum dots in the Schottky diode is calculated using one-dimensional Poisson-Schr\"{o}diner solver, and verified by the bias voltage at which the photocurrent signal changed sign for non-resonant laser excitation~\cite{Mar2011}. A positive bias voltage of 0.75 V is required to achieve zero total electric field in this structure.

\textbf{Quantum Confined Stark Effects.} Figure 2(a) shows the PL spectra of X$^{-}$ and X$^{2-}$ as a function of bias voltage at different magnetic fields applied along the growth direction, from base to apex of quantum dots. This direction is denoted as $z$ direction in this work. It can be seen that both X$^{-}$ and X$^{2-}$ peaks split in the presence of the magnetic field due to Zeeman effect, and the average energy of the splitted peaks shifts to high energy under the diamagnetic effect. The relative intensity of X$^{-}$ increases with increasing magnetic field because the probability for the in-plane ($xy$ plane) electrons being captured by quantum dot is reduced by the applied magnetic field in $z$ direction as discussed before \cite{Tang2014}.

For each magnetic field, the peak energy of charged excitons shifts as a function of external bias by the quantum-confined Stark effect in quantum dots~\cite{Mar2011,Xu20081,Warburton2002}. A striking feature is that the peak shifts to the high energy side at zero magnetic field, but the splitted peaks shift to the low energy side at high magnetic fields, for example, at 9 T. Similar features have been observed in other quantum dots from the same wafer as shown in Supplementary Figure S2 as another example. Owing to the built-in dipole in quantum dots, an asymmetric Stark shift in the presence of an applied electric field should be observed~\cite{Barker2000}. The observed asymmetric Stark shift gives the sign of built-in dipole moment, which indicates the spatial distribution of electron-hole wave functions in a single quantum dot. The sign of built-in dipole has been investigated over ten years both experimentally~\cite{Fry2000} and theoretically~\cite{Sheng2003,Barker2000} in single quantum dots with different shapes. By our definition, a positive built-in dipole moment shows that the hole wave function localizes towards the base of quantum dots under that of electrons.

Figure 2(b) shows the Stark shift of X$^{2-}$ as a function of total electric field with $B$= 0 T. The solid red line is the fitted result using the parabolic equation, which gives values for $E_{0}$, $p$ and $\beta$ for X$^{2-}$ as shown in the inset in Figure 2(b). A positive $p$ at zero electric field is achieved, which means that the center of effective mass of hole wave function is localized under that of electron in the quantum dot. This electron-hole alignment is consistent with results for pure InAs quantum dots with an ideal or truncated pyramidal shape~\cite{Finley2004}. However at 7 T, negative values of $p$ are obtained for each branch of the Zeeman splitting, as shown in the inset in Figure 2(c). More data of the Stark shift of the quantum dot at 4T and 9T are shown in the Supplementary Figure S3.

\textbf{Inverting Permanent Dipole.} Figure 3(a) shows the permanent dipole moment of the quantum dot as a function of applied magnetic field. For doubly charged state X$^{2-}$, the dipole $p/e$ is about 0.018 nm at zero magnetic field, then decreases with applying magnetic field over 4 T, and becomes negative when the applied magnetic field is over 7 T. Similar negative permanent dipole moment can be observed for X$^{-}$ with magnetic fields of 8 T and 9 T (as shown in the Figure 3(a) with solid triangles). This reveals that the permanent dipole is inverted in the quantum dot with applied magnetic fields.

With an eight-band \textbf{k$\cdot$p} method, the calculated confining potential for heavy holes increases linearly from base to apex for pure InAs quantum dots in truncated shape, which deduces that the hole wave functions are located towards the base~\cite{Sheng2001}. For electrons, however, the potential is more uniform, resulting in electron wave function occupying the center of the quantum dot. This gives a positive dipole moment, as observed in the quantum dot without a magnetic field. When a magnetic field is applied in $z$ direction, it provides an extra confinement for both electrons and heavy holes in $xy$ plane by cyclotron energy in quantum dots~\cite{Halonen1992,Babinski2006}. The confinement by magnetic field can be given by the magnetic length $l_{B}=\sqrt{\hbar/eB}$, where $\hbar$ is the plank constant, $e$ is the element charge, and $B$ is the applied magnetic field. For single quantum dots, the wave functions have been mapped with a diameter in $xy$ plane over 10 nm~\cite{Barker2000,Maltezopoulos2003,Maruccio2007}. This scale is comparable to the magnetic length 12.83 nm for single particle at 4 T, which explains that the permanent dipole begins to change from 4 T. Due to relatively uniform confinement of electrons, the center-of-mass of the electron wave functions does not change much in $z$ direction with magnetic field. At the base, the magnetic field shrinks the hole wave function in the plane. On account of the confining potential change, the shrinkage pushes the hole wave function center towards the apex of the quantum dot, resulting in an inverted electron-hole alignment. The wave functions at different magnetic fields applied along the base-apex direction are sketched in the inset in Figure 3(a).

\textbf{Discussion}

Next we discuss why this phenomenon can be observed here. One reason is that the permanent dipole is small, which makes the dipole inverting easily to observe with magnetic field in $z$ direction. The small permanent dipole moment could be due to the flat shape of quantum dots, as shown in the inset in the bottom panel of Figure 1(a). To address single quantum dots with relatively large metal apertures, we need a low dot density ($<$1 dot/$\mu$m$^2$) wafer. A technique of non-rotating substrate during the quantum dot growth was used to achieve a graded quantum dot density on a two-inch wafer. In the low density region grown by this method~\cite{Tanaka1999,Sun2004}, the typical sizes of quantum dots are about 25-50 nm in diameter of the base, 2-5 nm in height, which are similar to our quantum dots (as shown in the inset in Figure 1(a)). This size means that the quantum dots in low dot density region are relatively flat in shape, which gives a small permanent dipole in $z$ direction. The permanent dipole ($p/e$) around 0.018 nm without magnetic field in our sample is one order of magnitude less than the thick quantum dots \cite{Fry2000}. The polarizability is also very sensitive to the height of the quantum dots, the higher quantum dot the larger polarizability \cite{Barker2000}. Here the polarizability ($\beta/e$) with a value around or even less than 1 nm$^2$/V is much smaller than other results reported (as shown in Figure 3(b)), which further confirms that the height of the quantum dots is small.

The charging states (X$^{-}$ and X$^{2-}$) instead of neutral states (X$^{0}$) investigated in this work have smaller permanent dipole moments as well~\cite{Mar2011,Finley2004}. The centers of effective mass of electrons and holes move closer with additional electrons added to the dots for X$^{-}$ and X$^{2-}$~\cite{Mar2011}, which is similar to positively charged exciton states (X$^{+}$) as reported in Ref.38. The neutral excitons (X$^{0}$) are out of this voltage range with the excitation power, thereby the Stark effect of X$^{0}$ is not discussed in this work \cite{Tang2014}.

In summary, we have demonstrated that the charge wave functions can be controlled longitudinally with applying a magnetic field in $z$ direction to a pyramidal quantum dots. This is confirmed by tuning and inverting the permanent dipole of the negatively charged excitons in the quantum dots using a bias-controlled photoluminescent spectroscopy. The magnetic field induced confinement pushes the hole wave function towards the apex of quantum dot, resulting center of hole wave function moving to the top of electrons at high magnetic fields. In addition to manipulating the wave function transversely with magnetic field, the longitudinal wave function control provides an alternative way to manipulate the charge distribution. Further more, precisely tuning the electron-hole alignment could also be very useful to control the electron-hole interaction for solid-state quantum information processing, such as, entanglement generation~\cite{Stevenson2006}.


\textbf{Methods}

\textbf{Device Structure.} The sample was grown by molecular beam epitaxy and a schematic
structure is shown in the Supplementary Figure S1. On an [100]-oriented intrinsic GaAs
substrate, a distributed Bragg reflector of 13-pairs of
Al$_{0.94}$Ga$_{0.06}$As/GaAs (67/71 nm) was grown firstly, followed
by a 200 nm intrinsic GaAs buffer layer. The main device structure
consists: a Si $\delta$-doped GaAs layer with a doping density
N$_{d}$ = 5$\times$ 10$^{12}$ cm$^{-2}$ forming a two dimensional
electron gas (2DEG) region, a 50 nm intrinsic GaAs tunneling layer,
an InAs quantum dots layer with a dot density less than $1\times$10$^{9}$
cm$^{-2}$ with a growth temperature at 520 $^{o}$C, and a 200 nm intrinsic GaAs barrier layer grown on the
top.

\textbf{Fabrications.} To fabricate Schottky diode devices, mesa isolation trenches
were firstly patterned on the wafer by optical lithography. A layer
of AuGeNi was evaporated and alloy annealed at 420 $^{o}$C to form
ohmic contact with the Si $\delta$-doped GaAs layer in n-type
region. At the active regime, a semitransparent Ti with a thickness
of 10 nm was evaporated as Schottky contact, followed by an Al mask
with different apertures of 1-2 $\mu$m in diameter for addressing
single quantum dots. Finally, Cr/Au bond pads were evaporated on top
of the ohmic contact and the Al masks for connecting the electrical
wires.

\textbf{Optical Measurement.} The device was mounted on an $\emph{xyz}$ piezoelectric stage with
40 $\mu$m travel along each axis, then stage was placed in a helium
gas exchange cryostat equipped with a superconducting magnet. Measurements were carried out at 4.2 K. A magnetic field up to 9 T
was applied in $z$ direction, and the bias voltage was
supplied by a DC voltage source. To perform the micro-PL measurement
with a confocal microscopy system, a semiconductor laser with a
wavelength of 650 nm was used as pumping source and was focused on
one of the apertures by a microscope objective with a large
numerical aperture  of NA=0.83. The PL from single quantum dots was
collected with the same objective and dispersed through a 0.55 m
spectrometer, and the spectrum was detected with a liquid nitrogen
cooled charge coupled device camera with a spectral resolution of 60
$\mu$eV.

\textbf{References}
\bibliographystyle{naturemag}

\begin{addendum}
 \item [Acknowledgments] This work was supported by the National Basic Research Program of
China under Grant No. 2013CB328706 and 2014CB921003; the National
Natural Science Foundation of China under Grant No. 91436101, 11174356 and
61275060; the Strategic Priority Research Program of the Chinese
Academy of Sciences under Grant No. XDB07030200; and the Hundred Talents
Program of the Chinese Academy of Sciences. We
thank Andrew Ramsay, Jonathan Mar and Mete
Atat\"{u}re for very helpful discussions.
\item[Author contributions] X.X. and K. J. conceived and designed the project; S. C., J. T., Y. G., Y. S., K. Q. and X. X. performed the experiments; M. H., J. S. and L. G. did the TEM characterization; D. W. provided the wafers; S. C., J. T., Y. Z., W. S. and X. X. analysed the data; X. X. wrote the paper with contributions from other authors.

    \item [Additional information] Supplementary information accompanies this paper at http://www.nature.com/
scientificreports
 \item[Competing Interests] The authors declare that they have no
competing financial interests.

\end{addendum}

\newpage
\begin{figure*}
\centering
\epsfig{file=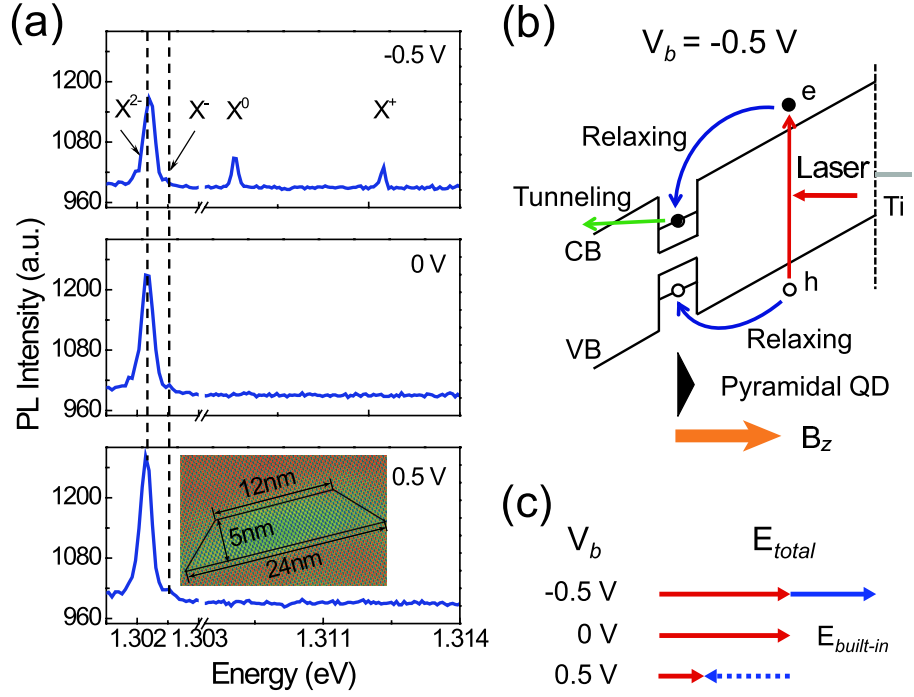,width=12cm,keepaspectratio}

\caption{\label{fig:figure1} (a) PL spectra at the bias voltages of -0.5 V,
0 V and +0.5 V from top to bottom panels. The PL emission lines of different charging states~$X^{2-}$,~$X^{-}$,~$X^{0}$ and $X^{+}$ are labeled in the
figure. The dotted lines are used to guide the eyes. Inset: A high-resolution cross section image of a single quantum dot by transmission electron microscope. (b) Band profiles of the
\emph{n-i-}Schottky diode structure under bias voltages (V$_{b}$) of -0.5 V. The apex of the pyramidal quantum dot orientates towards the Schottky contact as shown in the inset. The magnetic field  is applied along same direction as well ($z$ direction). (c) The total electric field (E$_{total}$, solid arrows) under different bias with considering the built-in electric field (E$_{built-in}$, solid red arrows). The dashed arrow marks the positive bias induced electric field.}

\end{figure*}

\begin{figure*}
\centering
\epsfig{file=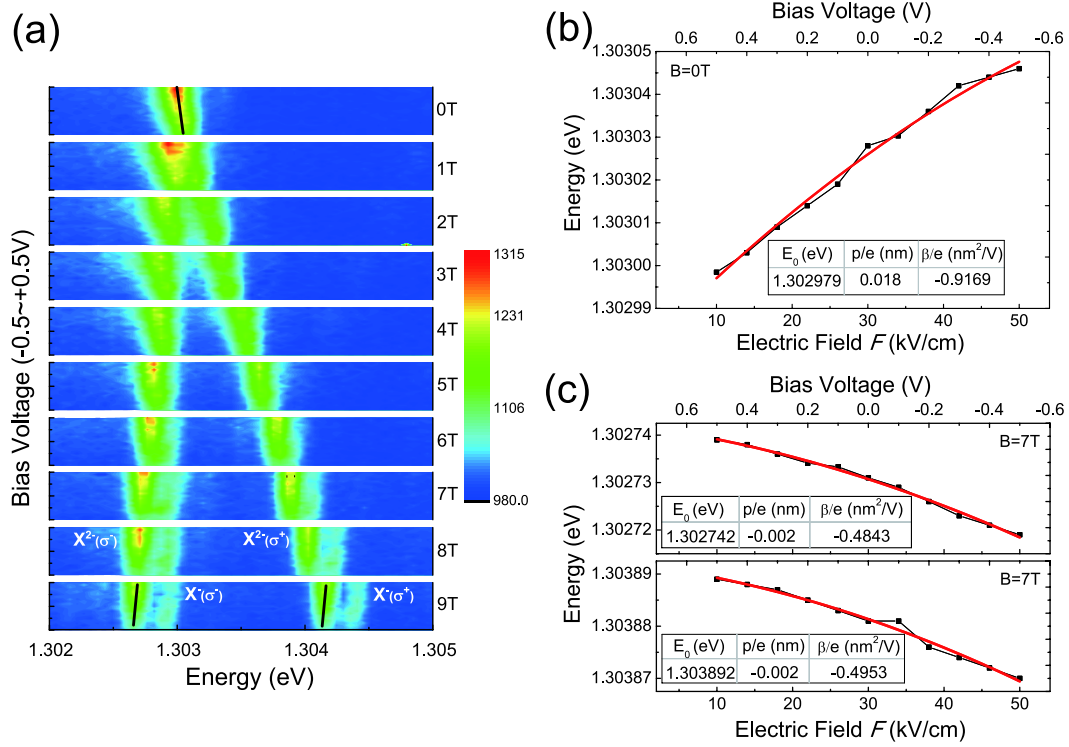,width=14cm,keepaspectratio}

\caption{\label{fig:figure2} (a) The contour plots of the PL spectra of $X^{2-}$ and $X^{-}$ as a function of bias voltage from -0.5 V to +0.5V at different magnetic fields. Due to the Zeeman splitting, four peaks of $X^{2-}$ and $X^{-}$ at high magnetic fields can be observed, as marked in the Figure. The solid black lines are used to guide the eyes for the Stark shifts. (b) Transition energies (black square) of $X^{2-}$ as a function of E$_{total}$ across the quantum dot. The solid red line shows the fitted result and the fitted parameters are shown in the inset. (c) Transition energies (black square) of the two branches of $X^{2-}$ at 7 T as a function of E$_{total}$. The solid red lines are the fitted results of the Stark effect. }

\end{figure*}

\begin{figure*}
\centering
\epsfig{file=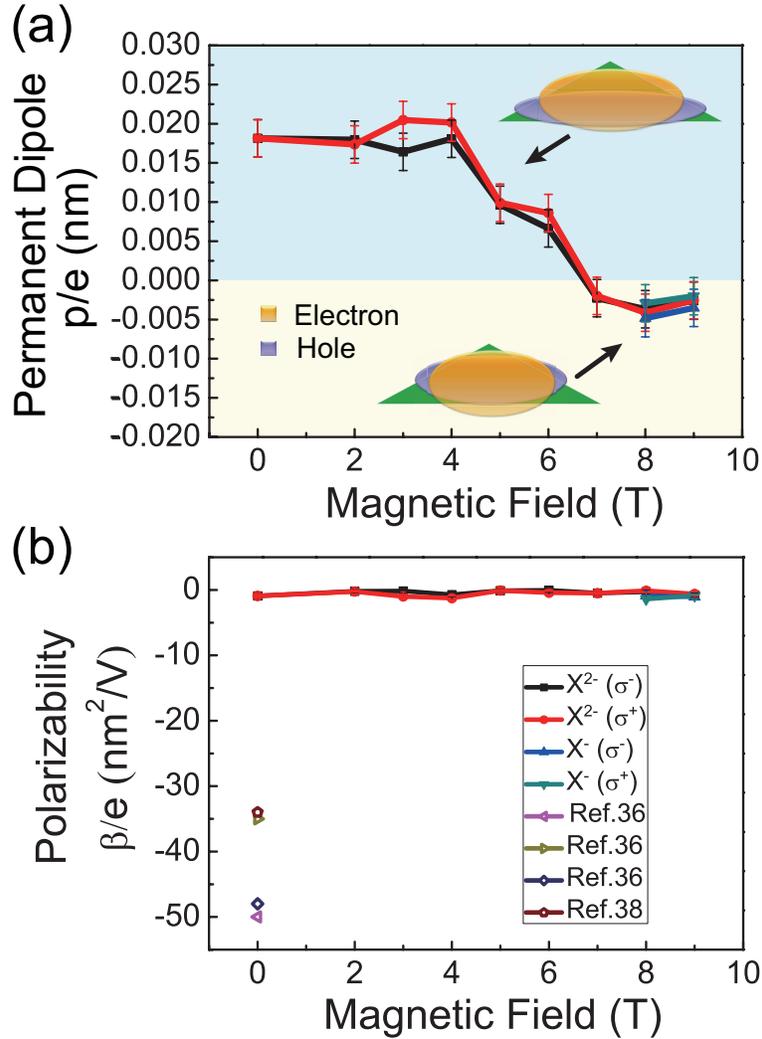,width=10cm,keepaspectratio}

\caption{\label{fig:figure3}(a) The permanent dipole (p/e) as a function of magnetic field. It can be seen that the sign of dipole moment inverts with increasing magnetic field. The electron/hole wave functions are sketched as shown in the inset. Without magnetic field, the electron/hole wave functions are confined by the quantum dot. Due to the smaller confinement of the base in the quantum dot, the heavy hole wave function locates towards the base, which gives a positive permanent dipole. With increasing magnetic field to 7 T, the magnetic field shrinks both electron and hole wave functions. For electron wave function, the center of effective mass does not change very much although the wave function spreads smaller in $xy$ plane. But for holes, the compressed wave function also pushes the center of the effective mass towards the apex of the quantum dot. (b) The polarizability of $X^{2-}$ and $X^{-}$ as a function of magnetic field. The empty symbols are data from Ref.36 and 38 for comparison.}

\end{figure*}

\end{document}